\documentclass[aps,prc,preprint,showpacs,floatfix]{revtex4-1}  
\usepackage{graphicx}
\usepackage{amsmath}

\newcommand {\nc} {\newcommand}
\nc {\beq} {\begin{eqnarray}} \nc {\eol} {\nonumber \\} \nc {\eeq}
{\end{eqnarray}} \nc {\eeqn} [1] {\label{#1} \end{eqnarray}} \nc
{\eoln} [1] {\label{#1} \\} \nc {\ve} [1] {\mbox{\boldmath $#1$}}
\nc {\rref} [1] {(\ref{#1})} \nc {\Eq} [1] {Eq.~(\ref{#1})} \nc
{\re} [1] {Ref.~\cite{#1}} \nc {\dem} {\mbox{$\frac{1}{2}$}} \nc
{\arrow} [2] {\mbox{$\mathop{\rightarrow}\limits_{#1 \rightarrow
#2}$}}
\begin{document}
\title{Detailed study of the astrophysical direct capture reaction $^{6}{\rm Li}(p, \gamma)^{7}{\rm
Be}$ in a potential model approach}

\author {E.M. Tursunov}
\email{tursune@inp.uz} \affiliation {Institute of Nuclear Physics,
Academy of Sciences, 100214, Ulugbek, Tashkent, Uzbekistan}
\affiliation {National University of Uzbekistan, 100174  Tashkent,
Uzbekistan}
\author {S.A. Turakulov}
\email{turakulov@inp.uz} \affiliation {Institute of Nuclear Physics,
Academy of Sciences, 100214, Ulugbek, Tashkent, Uzbekistan}
\author {K.I. Tursunmakhatov}
\email{tursunmahatovkahramon@gmail.com} \affiliation {Gulistan State
University,  120100, Gulistan, Uzbekistan}

\begin{abstract}
The astrophysical $S$ factor and reaction rates of the direct
capture process  $^{6}$Li(p,$\gamma)^{7}$Be are estimated within a
two-body single-channel potential model approach. Central potentials
of the Gaussian-form in the $^2P_{3/2}$ and  $^2P_{1/2}$  waves are
adjusted to reproduce the binding energies and the empirical values
of  the asymptotic normalization coefficients (ANC) for the
$^7$Be(3/2$^-$) ground and $^7$Be(1/2$^-$) excited bound states,
respectively. The parameters of the potential in the most important
$^2S_{1/2}$ scattering channel were fitted to reproduce the
empirical phase shifts from the literature and the low-energy
astrophysical $S$ factor of the LUNA collaboration.  The obtained
results for the  astrophysical $S$  factor  and the reaction rates
are in a very good agreement with available experimental data sets.
The numerical estimates reproduce not only the absolute values, but
also the energy  and temperature dependence of the $S$  factor   and
reaction rates of the LUNA collaboration, respectively.  The
estimated $^{7}{\rm Li/H}$ primordial abundance ratio $(4.67\pm 0.04
)\times 10^{-10}$  is well consistent with recent BBN  result of
$(4.72\pm 0.72) \times 10^{-10}$ after the Planck observation.

\end{abstract}


\maketitle

\section{Introduction}

\par One of the main problems of nuclear astrophysics is an estimation of the
present-day abundances of light elements \cite{cor2023,gro2023}. The
nuclear fusion processes with elements D, He and Li are important in
the Big Bang Nucleosynthesis (BBN) models. However, a large part of
present-day observed Li, Be and B abundances are due-to contribution
of galactic cosmic-ray spallation processes \cite{rolf1988}.

It  is established that less than half of of the present-day $^7$Li
abundance in the Solar system was produced during the BBN processes,
and a large part was synthesized in Stars \cite{fiel2020}. At the
same time, the present-day observed abundance of the $^6$Li element
is almost exclusively produced by cosmic-ray spallation processes of
heavy nuclei \cite{clay2003}. The Spite-plateau for the $^7$Li
primordial abundance, observed in old metal-poor halo Stars does not
exist for the $^6$Li element abundance \cite{spite1982,gnec2019}.

The main source of the primordial $^6$Li abundance in the BBN is
believed to be a nuclear direct capture reaction
d($\alpha,\gamma)^6$Li, while the reactions $^6$Li(p,$\alpha)^3$He
and $^{6}$Li(p,$\gamma)^{7}$Be are the most destruction channels
\cite{scha1977,noll1997}. A precise experimental results for the
astrophysical $S$ factor, reaction rates of the
d($\alpha,\gamma)^6$Li direct capture process and the primordial
abundance of the $^6$Li element obtained by the LUNA collaboration
in an underground facility \cite{LUNA2014,LUNA2017} recently have
been accurately described  within the three-body model
\cite{tur2016,tur2018a,tur2018b,tur2020,tur20b}. The theoretical
model reproduced not only the absolute values but also the energy
dependence of the astrophysical $S$ factor and the temperature
dependence of the reaction rates due-to correct treatment of the
isospin-mixing of about 0.5$\%$ in the final state. However, the
primordial abundance of the $^7$Li element is still a big challenge
for all the nuclear astrophysics community around the world, since
the astronomically observed abundance of this element is about 3
times less than the present-day BBN estimate \cite{LITHIUM2019}.

Thus, the $^6$Li/$^7$Li isotopic ratio is important to specify the
lithium production mechanisms either via cosmic-ray spallation
processes \cite{van2000} or via stellar evolution \cite{sack1999},
which modify the primordial $^7$Li abundance. For this purpose, an
accurate estimation of the astrophysical $S$ factor of the
$^{6}$Li(p,$\gamma)^{7}$Be direct capture process is necessary. The
cross section of this process influences many astrophysical
scenarios, including BBN and stellar evolution.

\par From the experimental side, very few measurements of the astrophysical $S$ factor of above reaction have been performed. The experiments were usually
carried out at higher energies due-to the Coulomb barrier problem,
so extrapolation to the low-energy region was necessary. In
addition, the enhancement due to the electron screening effect must
be taken into account \cite{ass1987}. The total uncertainty of the
old data set from Ref.\cite{swit1979} is significant ($\sim 15\%$)
that makes an extrapolation to the low astrophysical energy region
quite difficult \cite{NACRE99}. The most intriguing results have
been reported in Ref.~\cite{he2013} which found a resonance at
around 200 keV above the $p+^6$Li threshold with $J^{\pi}=1/2^+$,
which, however, was not seen in the data-set for the
$^3$He$(\alpha,\gamma)^7$Be capture reaction \cite{szucs2019}.
Recent direct measurement of the $^{6}$Li(p,$\gamma)^{7}$Be capture
reaction by the LUNA collaboration does not support the existence of
this resonance \cite{LUNA2020}.

\par The astrophysical $^{6}$Li(p,$\gamma)^{7}$Be direct capture reaction has also been  studied within various theoretical models, such as a potential model
\cite{xu2013}, a Gamow shell model \cite{dong2017}, cluster models
\cite{dub2022,gnec2019,arai2002} and R-matrix fits \cite{LUNA2020,
he2013,li2018}. The most successful description of the direct LUNA
data has been reached within the asymptotic normalization
coefficient (ANC) method \cite{kiss2021}. The empirical values of
the ANC for the  $^6$Li$+p \to^7$Be(3/2$^-$)  and   $^6$Li$+p
\to^7$Be(1/2$^-$)  virtual transitions to the  $^7$Be(3/2$^-$)
ground and $^7$Be(1/2$^-$) excited bound states have been derived
within the distorted wave Born approximation (DWBA) from the
analysis of the $^6$Li$(^3He,d)^7$Be transfer reaction.
Then on the base of deduced values of ANC the astrophysical $S$
factor of the $^{6}$Li(p,$\gamma)^{7}$Be direct capture reaction has
been estimated at low energies \cite{kiss2021}. As was proven in
many cases, indirect techniques like the ANC method or the Trojan
Horse method can give important clues to the understanding of BBN
(see, e.g., Ref.~\cite{piz2014} for a review). However, the above
work does not give a detailed information, how much is the
contribution of each entry channels to the capture process. In other
words, contributions of the partial $E$1-, $E$2- and
$M$1-astrophysical $S$ factors are not shown. On the other hand, the model does not probe the
most important experimental data for the reaction rates of the LUNA
collaboration \cite{LUNA2020}. These studies are important for
nuclear physics, since they allow to find the most realistic
$p+^6$Li potential parameters in each partial waves, both in the
bound and scattering channels. Only a model, which simultaneously
reproduces the absolute values and the energy dependence of the
astrophysical $S$ factor and temperature dependence of the reaction
rates, can be considered as fully realistic.

Very recently, a detailed study of the above
$^{6}$Li(p,$\gamma)^{7}$Be direct capture process at astrophysical
energies has been performed within the potential model
\cite{dub2022}. Various version of the potential model have been
suggested, however, no one of them describes the astrophysical $S$
factor and the reaction rates simultaneously. More precisely, the
temperature dependence of the reaction rates of the LUNA
collaboration \cite{LUNA2020} was not reproduced within that model.
Thus, a question, whether a potential model can describe the both
astrophysical $S$ factor and the reaction rates simultaneously,
remains open.

The aim of present work is a detailed study of the
$^{6}$Li(p,$\gamma)^{7}$Be astrophysical direct capture reaction at
low energies within a single-channel potential model where a channel
spin is fixed by $S$=1/2. Potential cluster models can
simultaneously describe the properties of bound states and
scattering data \cite{dub2022,gnec2019}. They can reproduce phase
shifts, binding energy, and an ANC. The importance of knowledge of
ANC in astrophysical processes was shown in particular in
Refs.~\cite{tur18,tur2021a,tur2021b}. A realistic potential model is
constructed based on the results of the ANC study of
Ref.~\cite{kiss2021} and the potential models of
Refs.~\cite{dub2011,dub2022}. In other words, a new potential model
will be able to reproduce the empirical ANC values for the
$^7$Be(3/2$^-$) ground and $^7$Be(1/2$^-$) excited bound states
deduced in Ref.~\cite{kiss2021} in addition to the experimental
bound state energies. The potential parameters in the $^2S_{1/2}$
wave are adjusted to reproduce the empirical phase-shifts
\cite{dub2011,dub2022} and the experimental astrophysical $S$ factor
of the LUNA collaboration \cite{LUNA2020}. Then the model will be
examined in reproducing the empirical reaction rates. The most
interest presents both absolute values and temperature dependence of
the theoretical reaction rates in comparison with the results of the
LUNA collaboration.

\par This article is organized as follows: a theoretical model will be
described in Section II, Section III is devoted to the analysis of
numerical results. Conclusions are given in the last section.

\section{The basic formalism of a two body single-channel model}

\subsection{Wave functions}

According to the single-channel potential model
\cite{tur18,tur2021a,tur2021b}, the initial and final states are
defined  by the factorized wave functions as
\begin{eqnarray}
\Psi_{lS}^{J}=\frac{u_E^{(lSJ)}(r)}{r}\left\{Y_{l}(\hat{r})\otimes\chi_{S}(\xi)
\right\}_{J M}
\end{eqnarray}
and
\begin{eqnarray}
\Psi_{l_f S'}^{J_f}
=\frac{u^{(l_fS'J_f)}(r)}{r}\left\{Y_{l_f}(\hat{r})\otimes\chi_{S'}(\xi)
\right\}_{J_f M_f},
\end{eqnarray}
respectively. The radial wave functions of the initial $p- ^6${\rm
Li} scattering states in the $^2S_{1/2}$, $^2P_{1/2}$, $^2P_{3/2}$,
$^2D_{3/2}$, $^2D_{5/2}$, $^2F_{5/2}$, $^2F_{7/2}$ partial waves are
described as solutions of the two-body Schr\"{o}dinger equation
\begin{align}
\left[-\frac{\hbar^2}{2\mu}\left(\frac{d^2}{dr^2}-\frac{l(l+1)}{r^2}\right)+V^{
lSJ}(r)\right] u_E^{(lSJ)}(r)= E u_E^{(lSJ)}(r),
\end{align}
where $\mu$ is the reduced mass of proton and $^6\rm Li$ nucleus,
${1}/{\mu}={1}/{m_1}+{1}/{m_2}$ , and $V^{lSJ}(r)$ is a two-body
potential in the partial wave with quantum numbers $l$ (orbital
angular momentum), $S$ (spin) and $J$ (total angular momentum). The
wave function $u^{(l_fS'J_f)}(r)$ of the final $^2P_{3/2}$ ground
and $^2P_{1/2}$ excited bound states are defined as a solution of
the bound-state Schr\"{o}dinger equation. The Schr\"{o}dinger
equation is solved using the high-accuracy Numerov algorithm.
At large distances the asymptotics of the scattering wave function
must satisfy the condition
\begin{equation}
 u_E^{(lSJ)}(r)=
\cos\delta_{lSJ}(E) F_l (\eta,kr) + \sin\delta_{lSJ}(E)
G_l(\eta,kr), \label{asymp}
\end{equation}
where $k$ is the wave number of the relative motion, $\eta$ is the
Zommerfeld parameter, $F_l$ and $G_l$ are regular and irregular
Coulomb functions, respectively, and $\delta_{lSJ}(E)$ is the phase
shift in the partial wave with quantum numbers $(l,S,J)$.

The $p-^6${\rm Li} two-body potential is chosen in the Gaussian form
as \cite{gnec2019,dub2022}:
\begin{equation}
 V^{lSJ}(r)=V_0 \exp(-\alpha_0 r^2)+V_c(r),
\label{pot}
 \end{equation}
where the Coulomb part of the potential is based on the point-like
charge distribution \cite{gnec2019,dub2022}.

\subsection{Cross sections of the radiative-capture process}

The cross section, the astrophysical $S$ factor and the reaction
rates are estimated using the accurate wave functions of the initial
and final states. The total cross section of the  radiative-capture
process is expressed as a sum of cross sections for each final
state \cite{tur2021a,tur2021b}:
\begin{eqnarray}
\sigma(E)=\sum_{J_f \lambda \Omega}\sigma_{J_f \lambda}(\Omega),
\end{eqnarray}
where $\Omega=$ $E$   (electric  transition) or $M$ (magnetic
transition), $\lambda$ is a multiplicity of the transition, $J_f$ is
the total angular momentum of the final state. For a particular
final state with total angular momentum $J_f$ and multiplicity
$\lambda$ we have \cite{tur2021a}
\begin{align}
 \sigma_{J_f \lambda}(\Omega) =& \sum_{J}\frac{(2J_f+1)} {\left
[S_1 \right]\left[S_2\right]} \frac{32 \pi^2 (\lambda+1)}{\hbar
\lambda \left( \left[ \lambda \right]!! \right)^2} k_{\gamma}^{2
\lambda+1} C^2(S) \sum_{l S}
 \frac{1}{ k_i^2 v_i}\mid
 \langle \Psi_{l_f S'}^{J_f}
\|M_\lambda^\Omega\| \Psi_{l S}^{J} \rangle \mid^2,
\end{align}
where $l$ and $l_{f}$ are the orbital momenta of the initial and
final states, respectively; $k_i$ and $v_i$ are the wave number and
speed of the $p - ^6${\rm Li} relative motion in the entrance
channel, respectively; $S_1$ and $S_2$ are spins of  $p$ and
$^6${\rm Li},  $S'=S=$1/2 due to the use of the single-channel
approximation. The $k_{\gamma}=E_\gamma / \hbar c$ is the wave
number of the photon corresponding to energy $E_\gamma=E_{\rm
th}+E$, where $E_{\rm th}$ is the threshold energy for the breakup
reaction $\gamma+^{7}$Be$\to^{6}$Li+p. Constant $C^2(S)$ is a
spectroscopic factor \cite{NACRE99}. Within the potential approach
where the bound and scattering properties (energies, phase shifts
and scattering length) are reproduced, a value of the spectroscopic
factor must be taken equal to 1 \cite{mukh2016}. We also use
short-hand notations $[S]=2S+1$ and $[\lambda]!!=(2\lambda+1)!!$.

The reduced matrix elements are estimated between the initial
$\Psi_{l S}^{J}$ and final $\Psi_{l_f S'}^{J_f}$ states. The
electric transition operator in the long-wavelength approximation
reads as
\begin{eqnarray}
M_{\lambda \mu}^{\rm E}=e \sum_{j=1}^{A}
Z_j{r'_j}^{\lambda}Y_{\lambda \mu}(\hat{r'}_j),
\end{eqnarray}
where $\vec {r'}_{j}= \vec{r}_j-\vec{R}_{cm}$ is the position of the
$j$th particle in the center of mass system.


The magnetic transition operator reads as \cite{tur18,tur2021a}
\begin{eqnarray}
M_{1 \mu}^{\rm M}&=& \sqrt{\frac {3}{4 \pi}} \sum_{j=1}^{A}
\left[\mu_N \frac{Z_j}{A_j}\hat{l}_{j \mu} + 2 \mu_j \hat{S}_{j \mu}
\right] \\ \nonumber
 & = & \sqrt{\frac {3}{4 \pi}}\left[\mu_N \left( \frac{A_2
Z_1}{A A_1}  \frac{A_1 Z_2}{A A_2} \right) \hat{l}_{r \mu}+
2(\mu_1\hat{S}_{1\mu}+\mu_2\hat{S}_{2\mu})\right],
\end{eqnarray}
where $\mu_N$ is the nuclear magneton, $\mu_j$ is the magnetic
moment and $\hat{l}_{j \mu}$ ($\mu=-1,0,+1$) is the projection of
the orbital angular momentum of $j$ th particle. The projection of
the orbital angular momentum of the relative motion is denoted as
$\hat{l}_{r \mu}$.

The explicit expressions for the reduced matrix elements of the
electric and magnetic transition operators  were given in
Ref.\cite{tur2021b}. In the above equations the spins of the
particles are $S_1=S_p=1/2$, $S_2=S(^6 Li)$=1 and magnetic momenta
are taken as $\mu_{p}=$2.792847$\mu_N$ and $\mu(^6\rm Li)$=0.822
$\mu_N$ for the first and second particles, respectively.

Finally, the astrophysical $S$ factor of the process is related to
the cross section as \cite{fow1975}
\begin{eqnarray}
S(E)=E \, \, \sigma(E) \exp(2 \pi \eta).
\end{eqnarray}

\section {Numerical results}

\subsection{Astrophysical $S$ factor of the $^{6}$Li(p,$\gamma)^{7}$Be reaction}

The wave functions of initial and final states are  found from the
numerical solution of the  Schr\"{o}dinger equation in the entrance
and exit channels with the $p-^6\rm {Li}$ central potentials of the
Gaussian-form as defined in Eq.(\ref{pot}) with the corresponding
point like Coulomb part. Hereafter everywhere the parameter values
$\hbar^2/2 u$ =20.9008 MeV fm$^2$, 1u=931.494 MeV, $m_p$=1.00727647
u , m($^6$Li)=6.01347746 u and $\hbar$c=197.327 MeV fm are used in
numerical calculations.

\begin{table}[htbp]
\centering \caption{Central $\textrm{V}_\textrm{M}$ potential
parameters for the p-$^6$Li interaction in different partial waves.}
\label{table1}
\begin{tabular}{c c c c c}
\hline$^{2S+1}L_J$ & $V_0$ (MeV)  & $\alpha_0$ (fm$^{-2}$) & $C_{LJ}$ (fm$^{-1/2})$ &$E_{FS}^{^7Be}$ (MeV)\\
\hline
$^2S_{1/2}$ & -52.0 & 0.297 & &-5.81 \\
$^2P_{3/2}$ & -76.6277 & 0.1750& 2.191 &-\\
$^2P_{1/2}$ & -74.8169 & 0.1731& 2.070 &-\\
$^2D_{3/2}$ & -86.0 & 0.094& &-6.95\\
$^2D_{5/2}$ & -88.0  & 0.094& &-7.75\\
$^2F_{5/2}$ & -111.6 & 0.10& &-\\
$^2F_{7/2}$ & -44.34  & 0.05& &-\\  \hline
\end{tabular}
\vspace*{0.5cm}
\end{table}

The  parameters of the Gaussian-form central potential  model
$\textrm{V}_\textrm{M}$ are presented in Table \ref{table1}. The
modified potential model $\textrm{V}_\textrm{M}$  differs from the
original model  $\textrm{V}_\textrm{D}$ of Ref.~\cite{dub2022}   in
the  $^2S_{1/2}$, $^2P_{1/2}$ , $^2P_{3/2}$, $^2D_{3/2}$ , and
$^2D_{5/2}$  partial waves, while keeping the  parameter values
unchanged in the  $F$-  wave scattering channels.

The parameters of the modified $\textrm{V}_\textrm{M}$ potential in
the  $^2P_{1/2}$ and $^2P_{3/2}$ partial waves are fitted to
reproduce the experimental energies $E(3/2^-)$=-5.6068 MeV and
$E(1/2^-)$=-5.1767  MeV and the empirical squared ANC values of
4.81$\pm$ 0.38 fm$^{-1}$   and  4.29$\pm$0.27 fm$^{-1}$  for the
$^7\rm{Be(3/2^-)}$  ground and  $^7\rm{Be(1/2^-)}$   excited bound
states, respectively.  The empirical values of  ANC  were  obtained
from the analysis of the experimental differential cross section of
the proton transfer reaction using modified DWBA approach in
Ref.~\cite{kiss2021}.

\begin{figure}[htbp]
\includegraphics[width=10 cm]{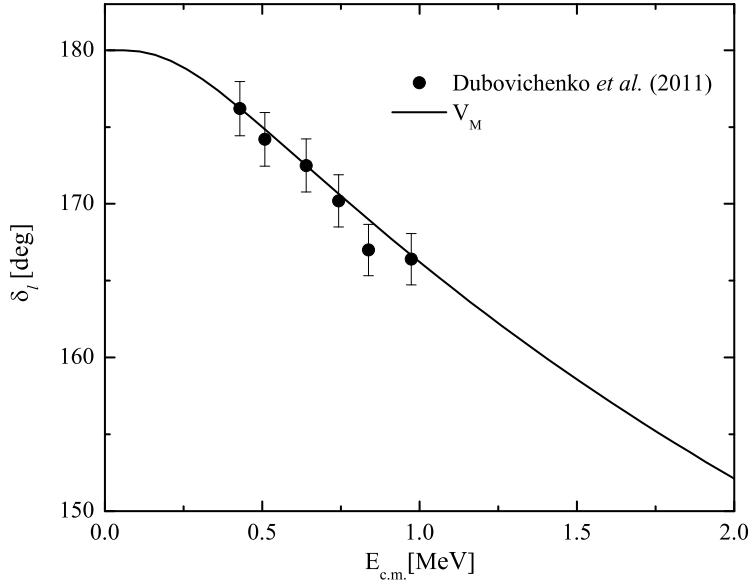} \caption{Phase shift in the $^2S_{1/2}$ partial wave of the $p-^6${\rm Li} scattering state within the potential model
$\textrm{V}_\textrm{M}$. The experimental phase shifts are taken
from Ref.~\cite{dub2011}} \label{fig1}
\end{figure}

The parameters of the modified $\textrm{V}_\textrm{M}$ potential  in
the  $^2S_{1/2}$ partial wave were fitted to reproduce the
experimental phase shifts   from Ref.~\cite{dub2011} and the direct
data of the LUNA collaboration for the astrophysical $S$ factor  of
the   $^{6}$Li(p,$\gamma)^{7}$Be  capture reaction at low energies
\cite{LUNA2020}. The calculated phase shifts for the p$-^6$Li
scattering in the $^2S_{1/2}$  partial wave using modified
$\textrm{V}_\textrm{M}$ potential  are presented in Fig.~\ref{fig1}
in comparison with the experimental data from \cite{dub2011}.  It
has been shown that $E$1 transitions to the final $P$ states play
the main role in the synthesis reaction  $^{6}$Li(p,$\gamma)^{7}$Be.
Therefore, the parameters of the  $\textrm{V}_\textrm{M}$ potential
in the  $S$-wave can be adjusted to the experimental values of the
astrophysical $S$ factor of the LUNA collaboration.  In other words,
from phase-equivalent potentials  one can choose  a model, which can
describe the experimental astrophysical $S$ factor  at low energies
at a satisfactory level. The parameters of the model in the $D$-wave
scattering channels are fitted to reproduce the astrophysical $S$
factor at larger energies. As was noted in the Introduction, the
main point here is a question, whether a potential  model,
constructed in such a way, will be  able to reproduce simultaneously
the  both absolute values and temperature dependence of the reaction
rates of the LUNA collaboration without any additional parameters.
This check is  realistic in a sense that a correct model must
describe the both  absolute values and temperature dependence of the
reaction rates, as was  demonstrated in the study of the direct
capture process   $\alpha$(d,$\gamma)^{6}$Li
\cite{tur2018b,tur2021b}.

Energies of the forbidden states in the $^2S_{1/2}$, $^2D_{3/2}$,
$^2D_{5/2}$ partial waves are given in the last column of   Table
\ref{table1}.  The Pauli forbidden state in the $S$-wave $p-^6$Li
relative motion corresponds to the orbital configuration [$s^7$],
while the Pauli forbidden states in the $D$-waves of the two-body
relative motion correspond to the  orbital scheme  [$s^5p^2$]
\cite{dub2022}.

\begin{figure}[htbp]
\includegraphics[width=10 cm]{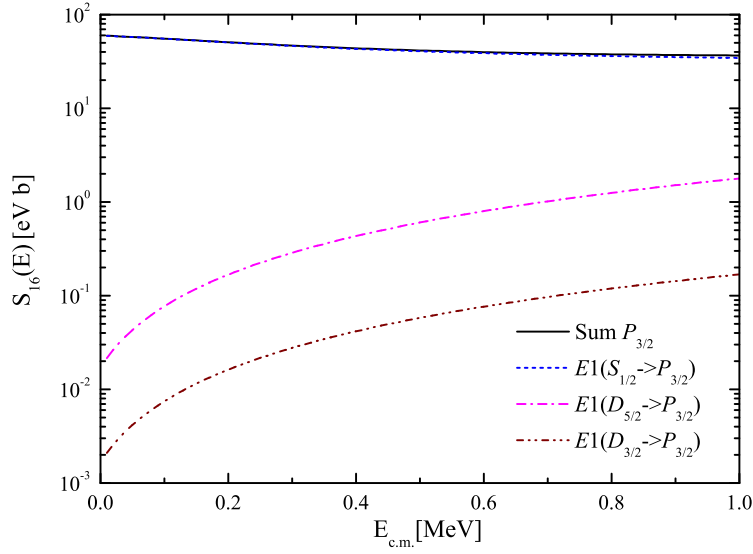}\caption{The  contributions of the $E$1 transition operator to the astrophysical $S$ factor  within the $\textrm{V}_\textrm{M}$ potential model from different initial scattering states to the final $^7\rm{Be(3/2^-)}$ ground state}
\label{fig2}
\end{figure}

The partial  contributions of the $E$1 transition to the
astrophysical $S$ factor for the $\textrm{V}_\textrm{M}$ potential
from different initial scattering states to the final
$^7\rm{Be(3/2^-)}$ ground state are shown in Fig.~\ref{fig2}. As is
seen from the figure, the dominant contribution corresponds to the
$^2S_{1/2}\to ^2P_{3/2}$  transition in the whole  energy region up
to 1 MeV.

\begin{figure}[htbp]
\includegraphics[width=10 cm]{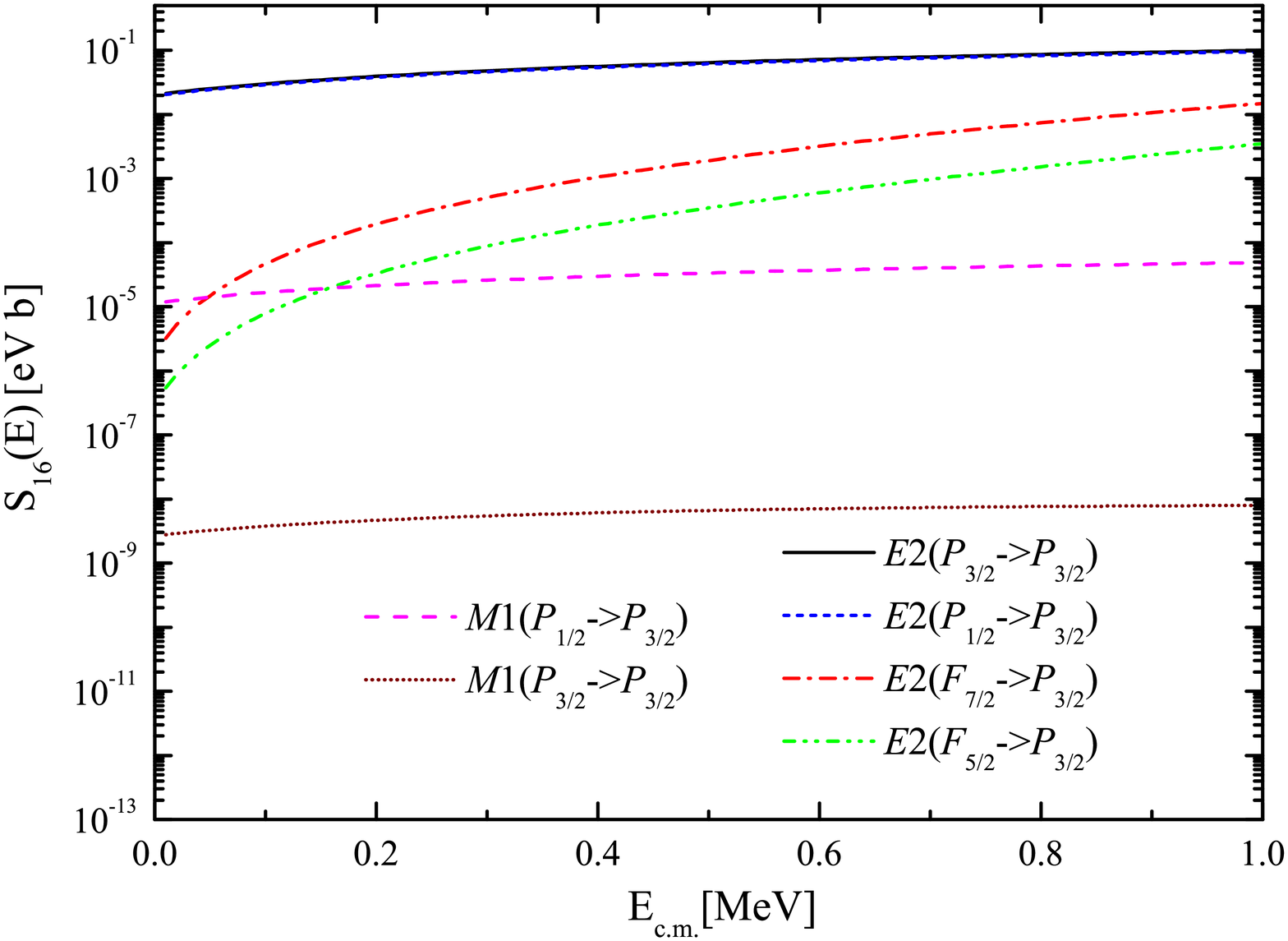}\caption{The partial  contributions of the $E$2 and $M1$ transition operators  to the astrophysical $S$ factor from different initial scattering states  to the final $^7\rm{Be(3/2^-)}$ ground state  within the $\textrm{V}_\textrm{M}$ potential model }
\label{fig3}
\end{figure}

The partial  contributions of the $E$2 and $M1$ transitions to the
astrophysical $S$ factor from different initial scattering states to
the final $^7\rm{Be(3/2^-)}$ ground state are shown in
Fig.~\ref{fig3}. As can be seen from the figure, these contributions
are negligible comparing to  the contribution of the $E1$ transition
in Fig.~ \ref{fig2}.  They  differ by more than three order of
magnitude.

\begin{figure}[htbp]
\includegraphics[width=10 cm]{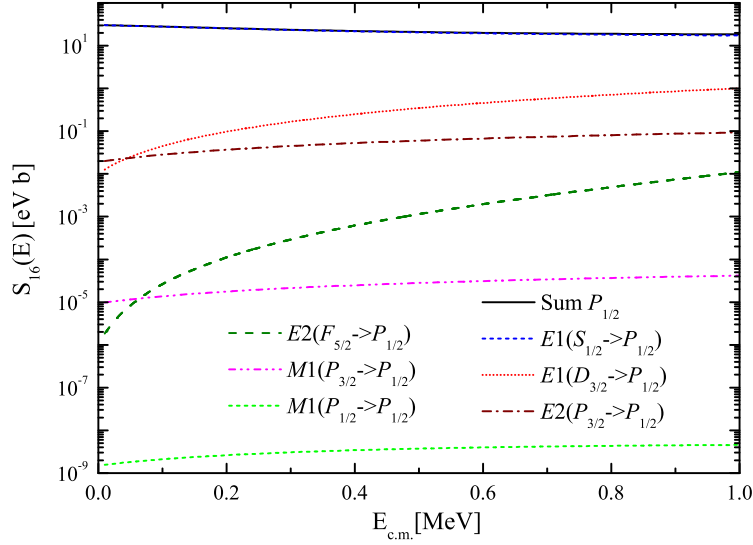}\caption{Contributions of  the  $E1$,  $E2$ and $M1$ transition operators  to the astrophysical $S$ factor from different initial scattering states  to the final $^7\rm{Be(1/2^-)}$  excited bound state         within the $\textrm{V}_\textrm{M}$ potential model}
\label{fig4}
\end{figure}

In Fig.~\ref{fig4} the partial  contributions of  all  $E1$,  $E2$
and $M1$ transitions  to the astrophysical $S$ factor from different
initial scattering states  to the final $^7\rm{Be(1/2^-)}$  excited
bound state are presented.   As can be noted here,  the most
important contribution comes from the $E1$ transition  $^2S_{1/2}\to
^2P_{1/2}$.

\begin{figure}[htbp]
\includegraphics[width=10 cm]{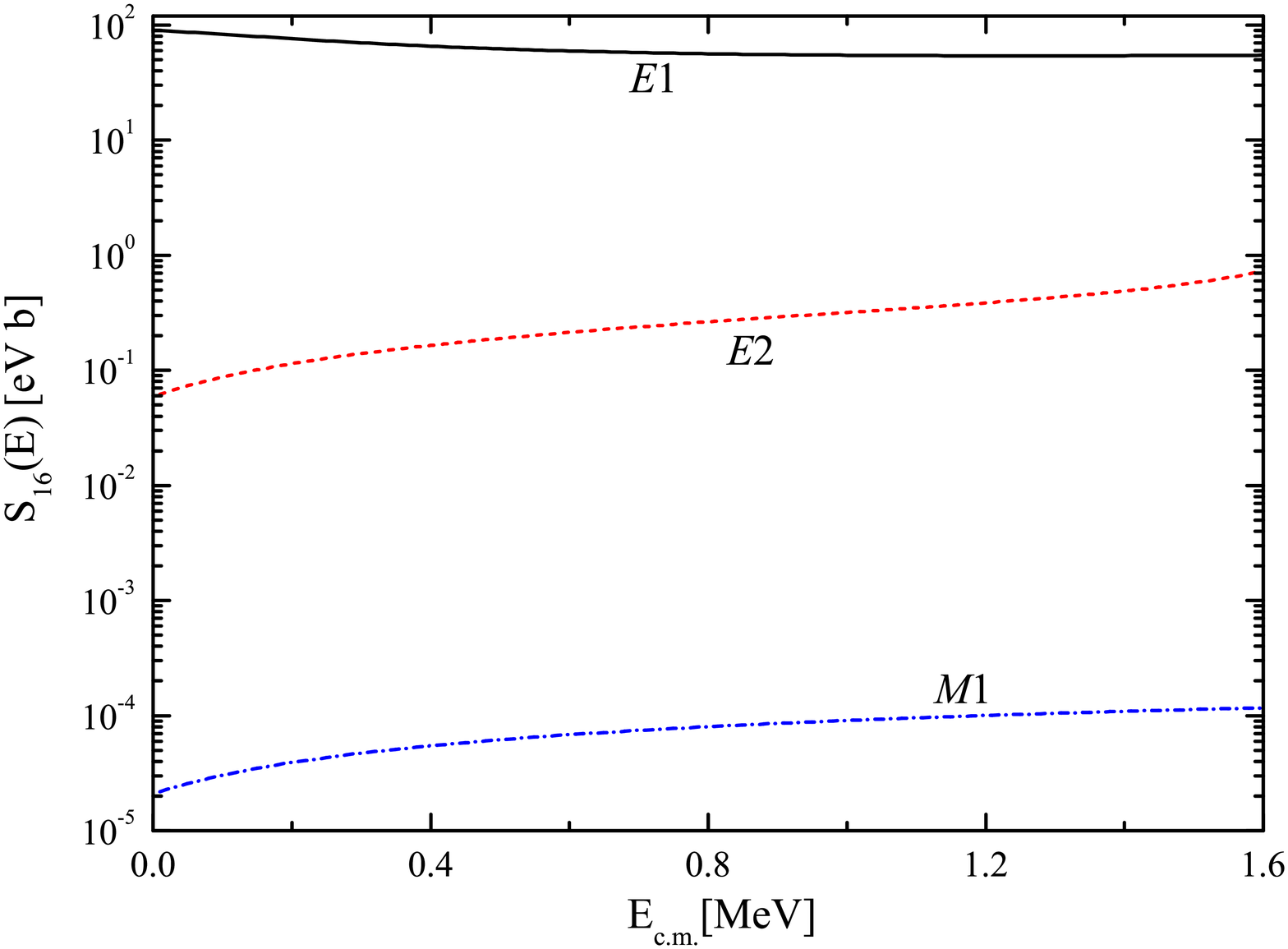}\caption{Contributions of E1, E2 and M1
transition operators to the astrophysical $S$ factor of the
$^{6}$Li(p,$\gamma)^{7}$Be synthesis process, calculated within the
potential  model $\textrm{V}_\textrm{M}$} \label{fig5}
\end{figure}

Figure~\ref{fig5} shows a comparison of the contributions of the
$E$1, $E$2 and $M$1 transitions to the astrophysical $S$ factor of
the $^{6}$Li(p,$\gamma)^{7}$Be synthesis process. As can be seen
from the figure, the $E$1 transition yields a dominant contribution
in the entire energy range up to 1.0 MeV. The contributions of $E$2
and $M$1 transitions  are much more suppressed. Even the
contribution of the  $E$2 transition is less than the contribution
from the $E$1 transition by three order of magnitude at low energy
region close to 0 and by two order of magnitude around the energy
value $E$=1 MeV.

\begin{figure}[htbp]
\includegraphics[width=10 cm]{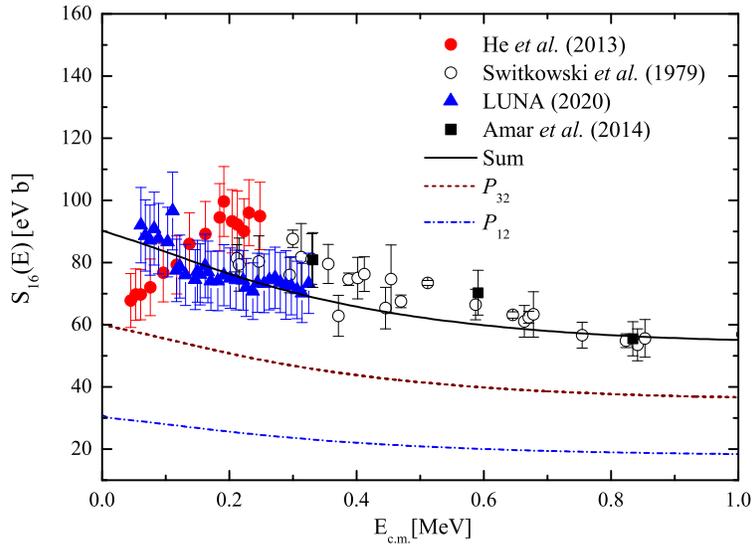}\caption{Astrophysical $S$ factor
of the radiative direct capture $^{6}$Li(p,$\gamma)^{7}$Be reaction,
calculated with the potential $\textrm{V}_\textrm{M}$ in comparison
with the experimental data from
Refs.~\cite{swit1979,he2013,LUNA2020,amar2014}} \label{fig6}
\end{figure}

A comparison of the theoretical astrophysical $S$ factor of the
radiative direct capture $^{6}$Li(p,$\gamma)^{7}$Be reaction with
the experimental data sets from Refs.~ \cite{swit1979,he2013,
LUNA2020,amar2014} are shown in Fig.~\ref{fig6}.  As can be seen
from Fig.~\ref{fig6}, the calculated astrophysical $S$ factor with
the  $\textrm{V}_\textrm{M}$ potential  model is in good agreement
with the direct experimental data of the LUNA Collaboration
\cite{LUNA2020} at low energies.  It also gives an overall good
description  of  other experimental data  sets  at energy range below 1.0 MeV except  the data set
from Ref.~\cite{he2013}.

In Table ~\ref{tab1a}  we give the calculated values of  the
astrophysical $S$ factor of the direct $^{6}$Li(p,$\gamma)^{7}$Be
capture reaction separately for the  ground $^7$Be($3/2^{-}$) and
first excited $^7$Be($1/2^{-}$)  bound states and their sum,  at
several energies, including  the zero and Gamow energy of $E$=15.1
keV.   The zero-energy  astrophysical $S$ factor was determined by
using the asymptotic expansion method of Ref. \cite{baye00}.
\begin{table}[htbp]
\centering \caption{The calculated values of astrophysical $S$
factors for the ground $^7$Be ($3/2^{-}$), first excited
$^7$Be($1/2^{-}$) ($E^*=0.429$ MeV) bound states and their sum at
energies $E$=0, 15.1 keV and 25 keV.} \label{table2}
 \begin{tabular}{c c c c}
\hline E$^{\rm{*}}$ & S(0)   & S(15.1 keV) & S(25 keV) \\
      \rm{[MeV]} & \rm{[eV b]}  & \rm{[eV b]} & \rm{[eV b]} \\
\hline
0.0 & 60.0 & 59.32 & 48.67 \\
0.429 & 30.31 & 29.97& 24.54\\
Total & 90.31 & 89.29& 73.21\\
\hline
\end{tabular} \label{tab1a}
\vspace*{0.5cm}
\end{table}

\subsection{Reaction rates of $^{6}{\rm Li}(p,\gamma)^{7}{\rm Be}$
process and primordial abundance of the $^7$Li element}

\par In nuclear astrophysics one of the most important input quantity for the estimation of primordial abundances of chemical elements in the Big Bang model of the Universe is the rates  of the   basis nuclear reactions.  The reaction rate $N_{A}(\sigma v)$ is calculated on the basis of calculated  cross-section of the process with the help of following expression\cite{NACRE99,fow1984}
\begin{eqnarray}
N_{A}(\sigma v)=N_{A}
\frac{(8/\pi)^{1/2}}{\mu^{1/2}(k_{\text{B}}T)^{3/2}} \times \nonumber \\
\int^{\infty}_{0} \sigma(E) E \exp(-E/k_{\text{B}}T) d E,
\end{eqnarray}
where $k_{\text{B}}$ is the Boltzmann coefficient, $T$ is the
temperature, $N_{A}=6.0221\times10^{23}\, \text{mol}^{-1}$ is the
Avogadro number. The reduced mass $\mu$  is determined from the
reduced mass number $A=A_1 A_2/(A_1 + A_2)$ for the $p+^{6}{\rm Li}$
system. When a variable $k_{\text{B}}T$ is expressed in units of MeV
it is convenient to use a variable $T_9$ for the temperature in
units of $10^9$ K according to the equation
$k_{\text{B}}T=T_{9}/11.605$ MeV. In present calculations $T_9$
varies within the interval $0.001\leq T_{9} \leq 10$.

In these new variables the above integral for the reaction rates can
be expressed as:
\begin{eqnarray}
\label{rate}
 N_{A}(\sigma v)=3.7313 \times 10^{10}A^{-1/2}\,\, T_{9}^{-3/2} \times \nonumber \\
\int^{ \infty}_{0} \sigma(E) E \exp(-11.605E/T_{9}) d E.
\end{eqnarray}
\begin{table}[htbp]
\centering \caption{  Theoretical estimates of the direct $^{6}{\rm
Li}(p, \gamma)^{7}{\rm Be}$ capture reaction rate $N_{A}(\sigma v)$
($ \textrm{cm}^{3} \rm{mol}^{-1} \rm{s}^{-1}$) in the temperature
interval $10^{6}$ K $\leq T \leq 10^{10}$ K ($ 0.001\leq T_{9} \leq
10 $).}
\bigskip
\begin{tabular}{c c c c c c c c} \hline
~~~$T_{{\rm{9}}}$ ~~~~~~& ~~~~~~
$\textrm{V}_\textrm{M}$~~~~~~~~~&~~~
$T_{{\rm{9}}}$~~~ &~~~~~~$\textrm{V}_\textrm{M}$~~~~~~\\
\hline
0.001&$3.10\times 10^{-29}$ &0.14&$3.53\times 10^{-1}$\\
0.002&$6.75\times 10^{-22}$ &0.15&$4.86\times 10^{-1}$\\
0.003&$2.39\times 10^{-18}$ &0.16& $6.51\times 10^{-1}$\\
0.004&$4.09\times 10^{-16}$ &0.18&$1.09\times 10^{0}$ \\
0.005&$1.57\times 10^{-14}$ &0.20& $1.69\times 10^{0}$\\
0.006&$2.53\times 10^{-13}$ &0.25&$4.02\times 10^{0}$\\
0.007&$2.32\times 10^{-12}$ &0.30&$7.74\times 10^{0}$\\
0.008&$1.44\times 10^{-11}$ &0.35&$1.30\times 10^{1}$\\
0.009&$6.73\times 10^{-11}$ &0.40&$1.97\times 10^{1}$\\
0.010&$2.53\times 10^{-10}$ &0.45&$2.80\times 10^{1}$\\
0.011&$8.06\times 10^{-10}$ &0.5&$3.77\times 10^{1}$\\
0.012&$2.24\times 10^{-9}$ &0.6&$6.13\times 10^{1}$\\
0.013&$5.60\times 10^{-9}$ &0.7&$8.95\times 10^{1}$\\
0.014&$1.27\times 10^{-8}$ &0.8&$1.22\times 10^{2}$\\
0.015&$2.69\times 10^{-8}$ &0.9&$1.57\times 10^{2}$\\
0.016&$5.33\times 10^{-8}$ &1&$1.95\times 10^{2}$\\
0.018&$1.78\times 10^{-7}$ &1.25&$2.99\times 10^{2}$\\
0.020&$5.02\times 10^{-7}$ &1.5&$4.10\times 10^{2}$\\
0.025&$3.99\times 10^{-6}$ &1.75&$5.23\times 10^{2}$\\
0.030&$1.92\times 10^{-5}$ &2&$6.38\times 10^{2}$\\
0.040&$1.88\times 10^{-4}$ &2.5&$8.65\times 10^{2}$\\
0.050&$9.44\times 10^{-4}$ &3&$1.09\times 10^{3}$\\
0.060&$3.20\times 10^{-3}$ &3.5&$1.30\times 10^{3}$\\
0.070&$8.46\times 10^{-3}$ &4&$1.51\times 10^{3}$\\
0.080&$1.88\times 10^{-2}$ &5&$1.90\times 10^{3}$\\
0.090&$3.67\times 10^{-2}$ &6&$2.26\times 10^{3}$\\
0.10&$6.53\times 10^{-2}$  &7&$2.60\times 10^{3}$\\
0.11&$1.08\times 10^{-1}$  &8&$2.92\times 10^{3}$\\
0.12&$1.67\times 10^{-1}$  &9&$3.23\times 10^{3}$\\
0.13&$2.48\times 10^{-1}$  &10&$3.51\times 10^{3}$\\
\hline
\end{tabular}\label{tab2}
\end{table}

\begin{figure}[htbp]
\includegraphics[width=10 cm]{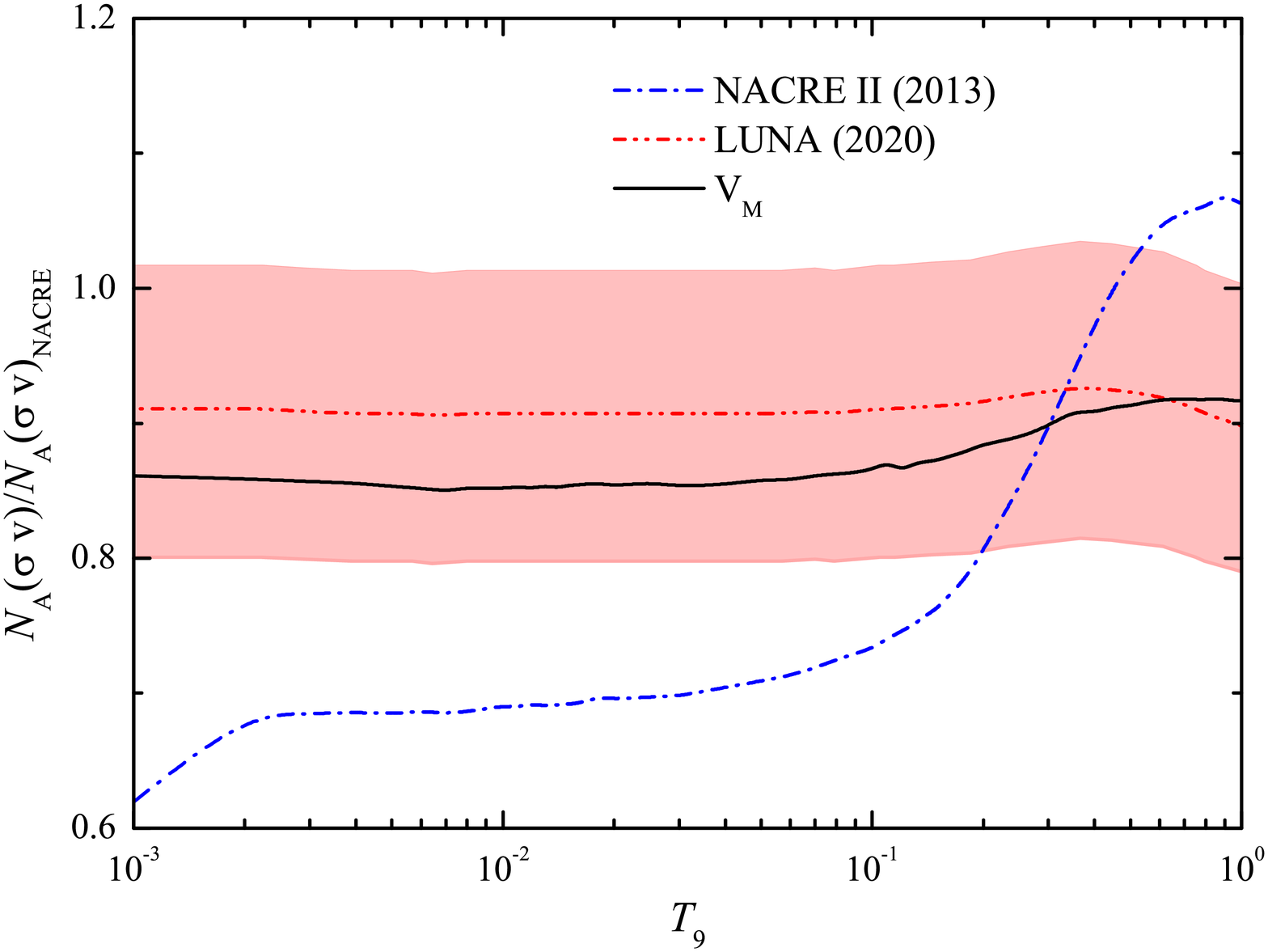}\caption{Comparison of the calculated
reaction rates in present potential model for the direct
$^{6}$Li(p,$\gamma)^{7}$Be capture process with the results of
Refs.~\cite{LUNA2020,xu2013}, normalized to the NACRE rate
\cite{NACRE99}} \label{fig7}
\end{figure}

The numerical values of the theoretically estimated  reaction rates
for the  $^{6}$Li(p,$\gamma)^{7}$Be direct capture process are given
in Table ~\ref{tab2}. A comparison of the calculated reaction rates
with the direct experimental data of the LUNA collaboration
\cite{LUNA2020} and the results of the NACRE II \cite{xu2013},
normalized to the NACRE rate \cite{NACRE99} are presented in
Fig.~\ref{fig7} in the temperature range from $T_9$=0.001 to
$T_9$=1. As can be seen from the figure, the new reaction rates
obtained in present work are  in very good agreement with the direct
data of the LUNA collaboration. In other words the present
theoretical model reproduces not only the absolute values of the
reaction rates of the direct experimental data of the LUNA
collaboration, but also the temperature dependence of the reaction
rates. As was noted in the Introduction, this point is the most
important result of the model, since we don't use any additional
parameters for reproducing the LUNA data for the reaction rates. Of
course, a success of the theoretical model is connected with a
realistic description of the $p+^{6}$Li interaction in important
scattering and bound state channels.

As mentioned above, for the estimation of the primordial abundance
of the $^7$Li element the theoretical reaction rate needs to be
approximated with the help of an analytical polynomial expression.
In our case the numerical results of reaction rate in
Table~\ref{tab2} are reproduced within 0.86 percent by using the
following analytical formula:
\begin{eqnarray}
N_{16}(\sigma v)=p_0 T_{9}^{-2/3} \exp ( -8.413 T_{9}^{-1/3}) \times \nonumber \\ ( 1 + p_1 T_9^{1/3} + p_2 T_{9}^{2/3}+p_3 T_{9}+p_4 T_{9}^{4/3} \nonumber \\
+ p_5 T_{9}^{-3/2}\exp (-5.634 T_9^{-1}),
\end{eqnarray}

The coefficients of the analytical approximation are given in
Table~\ref{tab3}.

\begin{table}[htbp]
\caption{Fitted values of the coefficients of analytical
approximation for the direct capture reaction $^{6}{\rm Li}(p,
\gamma)^{7}{\rm Be}$} {\begin{tabular}{@{}cccccccccc@{}} \toprule
 Model & $p_0$ & $p_1$ & $p_2$ & $p_3$ & $p_4$ & $p_5$\\
\colrule
$\textrm{V}_\textrm{M}$ & $1.032\times10^{6}$ & 0.369 & -0.929 & 0.492 & -0.075 &8.197 \\
\botrule
\end{tabular} \label{tab3}}
\end{table}

On the basis of the theoretical reaction rates and with the help of
the updated PArthENoPE 3.0 code \cite{pis2022} we have estimated a
contribution from the $^{6}{\rm Li}(p, \gamma)^{7}{\rm Be}$ direct
capture reaction to the primordial abundance of the $^7$Li element.
If we use the Planck 2018 data for the baryon density parameter
$\Omega_b h^2=0.02240\pm0.00010$(or $\eta_{10}=6.1322\pm0.0274$)
\cite{agh2020} and the new precision neutron life time $\tau_n=879.4
\pm 0.6$ s from Particle Data Group \cite{zyla2020}, for the
$^7$Li/H abundance ratio we have an estimate $(4.67 \pm 0.04)\times
10^{-10}$ within present potential model which is in good agreement
with the BBN result  $(4.72 \pm 0.72)\times 10^{-10}$ after the
Planck observation in Ref.~\cite{fiel2020}

\section{Conclusion}

Astrophysical direct nuclear capture reaction
$^6$Li+p$\to^7$Be+$\gamma$  was studied within the two-body single
channel potential model approach. The central $^6$Li-p potentials of
a Gaussian-form with the corresponding Coulomb part have been
examined. The parameters of the potential in the partial $^2P_{3/2}$
and $^2P_{1/2}$ waves  were fitted to reproduce the binding energies
and the empirical values of ANC for the $^7$Be(3/2$^-$) ground and
$^7$Be(1/2$^-$) excited bound states, respectively. The parameters
of the potential in the most important $^2S_{1/2}$ scattering wave
were fitted to reproduce the empirical phase shifts from the
literature and the low-energy astrophysical $S$ factor of the LUNA
collaboration. It is shown that the $E$1 transition from the initial
$S$-wave to the final $P$-waves yields a dominant contribution to
the astrophysical $S$ factor of the direct capture process. The
obtained results for the astrophysical $S$ factor of the
$^{6}$Li(p$,\gamma)^{7}$Be reaction is well consistent with the last
experimental data of the direct measurement of the LUNA
Collaboration and other  important experimental data sets from the
literature  in the energy region up to 1 MeV. The theoretical
reaction rates calculated within the model reproduce the both
absolute values and temperature dependence of the LUNA collaboration
results.

The estimated $^{7}{\rm Li/H}$ abundance ratio of $(4.67\pm 0.04
)\times 10^{-10}$ is in a very good agreement with the recent BBN
analysis of $(4.72\pm 0.72) \times 10^{-10}$ after the Planck
observation.

 \end{document}